\documentclass[11pt,a4paper]{article}
\usepackage[latin9]{inputenc}
\usepackage[active]{srcltx}
\usepackage{amsmath}
\usepackage{amsthm}
\usepackage{amssymb}

\makeatletter

\usepackage{amsfonts}
\usepackage{amsthm}
\usepackage{epstopdf}

\newif\iffigs\figstrue

\makeatother

\begin{document}
\begin{titlepage} \vspace{0.3cm}
 \vspace{1cm}

\begin{center}
\textsc{\Large{}{}{}{}{}{}\ \\[0pt] \vspace{0mm}
 About the Coleman Instantons in $D$ Dimensions }{\Large{}{}{}{}{}{}
}\\[0pt] 
\par\end{center}


\begin{center}
\vspace{35pt}
 \textsc{V. F. Mukhanov$^{~a,b}$ and A. S. Sorin$^{~c,d,e}$}\\[15pt] 
\par\end{center}

\begin{center}
{$^{a}$ Ludwig Maxmillian University, \\[0pt] Theresienstr. 37,
80333 Munich, Germany\\[0pt] }e-mail: \textit{\small{}{}{}{}{}{}mukhanov@physik.lmu.de}{\small{}{}{}{}{}\vspace{10pt}
 }{\small\par}
\par\end{center}


\begin{center}
{$^{b}$ {\small{}{}{}{}{}{}Korea Institute for Advanced Study\\[0pt]
Seoul, 02455, Korea}}\vspace{10pt}
\par\end{center}


\begin{center}
{$^{c}$ {\small{}{}{}{}{}{}Bogoliubov Laboratory of Theoretical
Physics\\[0pt] Joint Institute for Nuclear Research \\[0pt] 141980
Dubna, Moscow Region, Russia \\[0pt] }}e-mail: \textit{\small{}{}{}{}{}{}sorin@theor.jinr.ru}{\small{}{}{}{}{}\vspace{10pt}
 }{\small\par}
\par\end{center}


\begin{center}
{$^{d}$ {\small{}{}{}{}{}{}National Research Nuclear University
MEPhI\\[0pt] (Moscow Engineering Physics Institute),\\[0pt] Kashirskoe
Shosse 31, 115409 Moscow, Russia}}\vspace{10pt}
\par\end{center}

\begin{center}
{$^{e}$ {\small{}{}{}{}{}{}Dubna State University, \\[0pt]
141980 Dubna (Moscow region), Russia}}\vspace{10pt}
\par\end{center}


\vspace{1cm}

\begin{center}
\textbf{{Abstract} } 
\par\end{center}

We identify an infinite class of unbounded potentials for which the
Coleman instantons do not exist in $D$-dimensional spacetime. For
such potentials, the decay of a false vacuum is described by the new
instantons introduced in \cite{MRS1,MRS2,MS2}. For each spacetime
dimension $D$, we also construct the two different classes of potentials
for which the instanton equations are exactly solvable.

\newpage{}

\section{Obstacles to the existence of the Coleman instantons}

In 1978, S. Coleman, V. Glaser and A. Martin \cite{CGM} found a large
class of scalar field potentials $V\left(\varphi\right)$ in $D>2$
dimensions for which the Coleman instanton describing the false vacuum
decay exists and has a minimal action in the class of all solutions
with maximal $O\left(D\right)$ symmetry. Namely, they proved that
if for the continuous differentiable potential with a local minimum
at $\varphi=0$ there exist positive numbers $a,b,\alpha$ and $\beta$,
such that
\begin{equation}
\beta<\alpha<2D/(D-2),\label{eq:1b}
\end{equation}
and
\begin{equation}
V\left(\varphi\right)\geq a\left|\varphi\right|^{\beta}-b\left|\varphi\right|^{\alpha}\label{eq:2b}
\end{equation}
for all $\varphi$, then the Coleman instanton exists\footnote{Please note that in order to simplify the comparison of the results
with results of our previous work \cite{MS1}, we have changed some
notations compared to \cite{CGM}. In particular, we use the standard
notation for the scalar field potential $V\left(\varphi\right)$ instead
of $U\left(\Phi\right)$, have swapped $\alpha$ and $\beta$ compared
to \cite{CGM}, and denote the number of dimensions by $D$ instead
of $N$.}.

In the present work, we somewhat compliment the result of \cite{CGM}
and consider the problem of false vacuum decay in $D$-dimensional
scalar field theory under the assumption that the potential $V\text{\ensuremath{\left(\varphi\right)}}$
has a maximum at $\varphi=0$ and has no any local minima at positive
$\varphi$, is unbounded from below, and satisfies the inequality
opposite to (\ref{eq:2b}), i.e.
\begin{equation}
V\left(\varphi\right)<a\left|\varphi\right|^{\beta}-b\left|\varphi\right|^{\alpha}.\label{eq:3b}
\end{equation}
 We will prove that in this case the Coleman instanton, which is supposed
to describe the decay of the false vacuum at the absolute local minimum
at $\varphi_{f}<0$, does not exist regardless of the form of the
potential at negative $\varphi.$ Moreover, we will identify a broader
class of unbounded potentials for which the instantons with the Coleman
boundary conditions do not exist. 

For this purpose, we will consider the $O\left(D\right)$ Euclidean
solutions of the equation for the scalar field in $D$-dimensional
spacetime
\begin{equation}
\frac{\partial^{2}\varphi\left(t,\mathbf{x}\right)}{\partial t^{2}}-\Delta\varphi\left(t,\mathbf{x}\right)+\frac{dV(\varphi)}{d\varphi}=0.\,\label{eq:first}
\end{equation}
Having performed a Wick rotation, i.e. using the Eucledian time $\tau=it$,
and considering $O\left(D\right)$-invariant solutions, for which
$\varphi\left(\tau,\mathbf{x}\right)=\varphi\left(\varrho\right)$,
$\varrho\equiv\sqrt{\tau^{2}+\mathbf{x^{2}}}$,\textit{ }the equation
(\ref{eq:first}) reduces to the ordinary differential equation \cite{Coleman,CGM}
\begin{equation}
\ddot{\varphi}+\frac{D-1}{\varrho}\,\dot{\varphi}-V'=0,\label{eq:2}
\end{equation}
where $V'\equiv\frac{dV(\varphi)}{d\varphi}$ and $\dot{\varphi}\equiv\frac{d\varphi}{d\varrho}$.
The Coleman instanton satisfies the two boundary conditions, namely,
\begin{equation}
\varphi\left(\varrho\rightarrow\infty\right)=\varphi_{f}\,,\quad\dot{\varphi}\left(\varrho=0\right)=0\,.\label{eq:3}
\end{equation}
While the first condition must obviously be fulfilled, the second
condition is imposed to avoid a singularity at the ``center of the
bubble'', which would lead to a divergent infinite action. To prove
the absence of instantons with these boundary conditions for the unbounded
potentials satisfying (\ref{eq:3b}), we construct the following class
of nonlocal integrals for the equation (\ref{eq:2}):
\begin{align}
E\left(\alpha\right) & =\varrho^{\frac{2\,(\alpha\,(D-2)-2)}{\alpha+2}}\left(\varrho^{2}\,\left(\frac{1}{2}\,\dot{\varphi}^{2}-V\right)+\frac{2\,(D-1)}{\alpha+2}\,\varrho\,\varphi\,\dot{\varphi}-\frac{(D-1)\,\left(\alpha\,(D-2)-2\,D\right)}{(\alpha+2)^{2}}\,\varphi^{2}\right)\label{eq:5}\\
 & +\frac{2\,(D-1)}{\alpha+2}\,\int_{0}^{\varrho}d\overline{\varrho}\,\overline{\varrho}^{\frac{\alpha\,(2\,D-5)-6}{\alpha+2}}\left[\frac{\left(\alpha\,(D-2)-2\,D\right)\,\left(\alpha\,(D-2)-2\right)}{(\alpha+2)^{2}}\,\varphi^{2}+\overline{\varrho}^{2}\left(\alpha\,V-\varphi\,V'\right)\right]\,,~~~~~\nonumber 
\end{align}
which is parametrized by $\alpha$. To verify that it is indeed the
integral of motion for any value of $\alpha$, one can take its
derivative with respect to $\varrho$: 
\begin{equation}
\frac{d\,E\left(\alpha\right)}{d\varrho}=\varrho^{\frac{\alpha\,(2\,D-3)-2}{\alpha+2}}\left(\ddot{\varphi}(\varrho)+\frac{D-1}{\varrho}\,\dot{\varphi}(\varrho)-V'\right)\left(\varrho\,\dot{\varphi}+\frac{2\,(D-1)}{\alpha+2}\,\varphi\right)\,,\label{eq:6}
\end{equation}
which vanishes when equation (\ref{eq:2}) is satisfied. 

We note that integral (\ref{eq:5}) is invariant under the transformation:
$V\Rightarrow V+constant$. Therefore, without loss of generality
we can normalize the potential at the maximum such that $V\left(\varphi=0\right)=0$,
and assume that it has the deepest local minimum (false vacuum) at
$\varphi=\varphi_{f}<0$, has no any minima for $\varphi>0$ and
is unbounded from below, otherwise $V$ can be completely arbitrary.

Let us assume that the ``initial condition'' $\varphi\left(\varrho\rightarrow\infty\right)=\varphi_{f}$
is satisfied, and that there is the corresponding Coleman instanton
that allows the tunnelling from the false vacuum at $\varphi_{f}<0$
to $\varphi>0$. For this solution, the field $\varphi(\varrho)$
must vanish at a \textit{finite} value of $\varrho=\varrho_{0}$,
i.e. $\varphi\left(\varrho_{0}\right)=0$. Now, for $\alpha\geq\frac{2\,D}{D-2}$ we calcuate 
the values of the integral $E(\alpha)$ (\ref{eq:5}) at $\varrho=0$
\begin{equation}
E(\alpha)=0\,\label{Ealpha}
\end{equation}
and at $\varrho=\varrho_{0}$ 
\begin{align}
E\left(\alpha\right) & =\frac{1}{2}\,\varrho_{0}^{\frac{2\,\alpha\,(D-1)}{\alpha+2}}\,{\dot{\varphi}(\varrho_{0})}^{2}\label{eq:6a}\\
 & +\frac{2\,\left(D-1\right)}{\alpha+2}\,\int_{0}^{\varrho_{0}}d\overline{\varrho}\,\overline{\varrho}^{\frac{\alpha\,(2\,D-5)-6}{\alpha+2}}\left[\frac{\left(\alpha\,(D-2)-2\,D\right)\,\left(\alpha\,(D-2)-2\right)}{(\alpha+2)^{2}}\,\varphi^{2}+\overline{\varrho}^{2}\,\varphi^{\alpha+1}\,v'_{\alpha}\right]\,,\nonumber 
\end{align}
then compare the corresponding results, where the rescaled potential
$v_{\alpha}\left(\varphi\right)$ is defined as: 
\begin{equation}
v_{\alpha}\left(\varphi\right)\equiv-\varphi^{-\alpha}\,V\left(\varphi\right)\,.\label{eq:6b}
\end{equation}
If the Coleman instanton exists, the results of these two different
ways of calculating $E(\alpha)$ must be consistent, but they certainly
are not if 
\begin{equation}
v'_{\alpha}\equiv\frac{d\,v_{\alpha}\left(\varphi\right)}{d\varphi}>0\,.\label{eq:7}
\end{equation}
In fact, in this case the right-side of (\ref{eq:6a}) is obviously
positive, $E(\alpha)>0$, which contradicts to (\ref{Ealpha}) imposed
by the Coleman boundary condition $\dot{\varphi}\left(\varrho=0\right)=0$.
This implies that the field $\varphi(\varrho)$ cannot cross $\varphi=0$
to propagate towards $\varphi_{f}$, so it remains positive for all
finite values of $\varrho$, and hence the Coleman instanton does
not exist.

Finally, using the equations (\ref{eq:6b}) and (\ref{eq:7}), one
can conclude, that for any unbounded potential, which for positive
$\varphi$ can be represented as 
\begin{equation}
V(\varphi)=-\varphi^{\alpha}\int^{\varphi}d\bar{\varphi}\,v'_{\alpha}\left(\bar{\varphi}\right),\label{eq:8a}
\end{equation}
where
\begin{equation}
\alpha\geq\frac{2\,D}{D-2},\label{eq:9a}
\end{equation}
and
\begin{equation}
v'_{\alpha}\geq0,\label{eq:10a}
\end{equation}
the Coleman instanton does not describe the decay of the deepest false
vacuum at $\varphi_{f}<0$ regardless of the form of the potential
for negative values of $\varphi$. In particular, the potentials that
satisfy the condition (\ref{eq:3b}) and have no minima at positive
$\varphi$ belong to the class of the potentials (\ref{eq:8a}), so
that the Coleman instanton does not exist in this case. Other examples
of such potentials are presented in \cite{MS1} and we refer the reader
to that work. Therefore, the new regularized instantons \cite{MS2}
must necessarily be used for such potentials for which the false vacuum
is obviously unstable.

\section{Integrable potentials in $D$ dimensions}

Using the non-local integrals (\ref{eq:5}), we can determine two
classes of integrable potentials in $D$ - dimensions. For this purpose,
we first note that the first term under the integral in (\ref{eq:5})
vanishes if either $\alpha=2D/(D-2)$ or $\alpha=2/(D-2)$. In these
cases, the expression under the integral in (\ref{eq:5}) for the
potential $V\left(\varphi\right)=\lambda\,\varphi^{\alpha}$ vanishes
and the integral becomes local. Therefore, one can expect such potential
to be integrable. Let us start with the potential
\begin{equation}
V\left(\varphi\right)=\lambda\,\left(\varphi-\varphi_{0}\right)^{\frac{2D}{D-2}}+V_{0},\label{eq:11a}
\end{equation}
for which equation (\ref{eq:2}) becomes
\begin{equation}
\ddot{\varphi}+\frac{D-1}{\varrho}\,\dot{\varphi}-\lambda\,\frac{2\,D}{D-2}\left(\varphi-\varphi_{0}\right)^{\frac{D+2}{D-2}}=0.\label{eq:12a}
\end{equation}
If instead of $\varrho$ we introduce the new variable $\eta=\ln\varrho$,
for the rescaled field $\phi=e^{\frac{D-2}{2}\,\eta}\left(\varphi-\varphi_{0}\right)\,$,
we get the following equation 
\begin{equation}
\frac{d^{2}}{d\eta^{2}}\phi-\frac{\left(D-2\right)^{2}}{4}\,\phi-\lambda\,\frac{2
\,D}{D-2}\,\phi^{\frac{D+2}{D-2}}=0,\label{eq:13a}
\end{equation}
which has an obvious first integral.

For the potential
\begin{equation}
V\left(\varphi\right)=\lambda\,\left(\varphi-\varphi_{0}\right)^{\frac{2}{D-2}}+V_{0},\label{eq:14a}
\end{equation}
the equation
\begin{equation}
\ddot{\varphi}+\frac{D-1}{\varrho}\,\dot{\varphi}-\lambda\,\frac{2}{D-2}\,\left(\varphi-\varphi_{0}\right)^{\frac{4-D}{D-2}}=0,\label{eq:15a}
\end{equation}
reduces to an autonomous integrable equation, if one replaces $\varrho$
by $\eta=\varrho^{D-2}$ and then for the rescaled field $\phi=\eta\,\left(\varphi-\varphi_{0}\right)$
the above equation becomes
\begin{equation}
\frac{d^{2}}{d\eta^{2}}\phi-\lambda\,\frac{2}{\left(D-2\right)^{3}}\,\phi^{\frac{4-D}{D-2}}=0.\label{eq:16a}
\end{equation}
The parameters $\lambda$ and $V_{0}$ in the potentials (\ref{eq:11a})
and (\ref{eq:14a}) are arbitrary and can be negative or positive.
One could use these potentials to form piecewise exactly solvable bounded
and unbounded potentials with false and true vacua for arbitrary
dimensions $D>2$. In particular, the case of $D=4,$
where the exactly solvable potentials are $\varphi^{4}$ and $\varphi$,
was considered in detail in \cite{MRS1,MRS2}. For $D=3$ the exactly
solvable potentials are $\varphi^{6}$ and $\varphi^{2}$. 

\section{Conclusions}

We have considered the instantons for any $D>2$ and proved that for
the unbounded potentials which have no any local minima after
a maximum and do not satisfy the condition (\ref{eq:2b}) derived
in \cite{CGM}, the instantons satisfying the Coleman boundary conditions
(\ref{eq:3}) do not exist. In these cases, the instability of the
false vacuum is described by the new instantons \cite{MRS1,MRS2,MS1}.
Moreover, using the class of non-local integrals (\ref{eq:5}), we
have identified two classes of exactly solvable potentials in $D$
dimensions. These potentials can be used to build piecewise exactly
solvable potentials with the required properties in any number of
dimensions $D>2$. 

\bigskip{}

\textbf{Acknowledgments}

\bigskip{}

The work of V. M. was supported by the Germany Excellence Strategy---EXC-2111---Grant
No. 39081486.

The work of A. S. was supported in part by RFBR grant No. 20-02-00411.

\bigskip{}

\end{titlepage} 
\end{document}